\documentclass[11pt]{article}
\RequirePackage[T1]{fontenc}
\RequirePackage[utf8]{inputenc}
\RequirePackage[english]{babel}
\RequirePackage{lmodern}
\RequirePackage{microtype}
\RequirePackage{csquotes}
\RequirePackage[a4paper,top=2.35cm,bottom=2.35cm,left=2.55cm,right=2.55cm,footskip=1.15cm]{geometry}
\RequirePackage{amsmath,amssymb,amsthm,mathtools}
\RequirePackage{booktabs,tabularx,longtable,array,multirow}
\newcolumntype{Y}{>{\raggedright\arraybackslash}X}
\newcolumntype{L}[1]{>{\raggedright\arraybackslash}p{#1}}
\RequirePackage{graphicx}
\RequirePackage{enumitem}
\RequirePackage{xcolor}
\definecolor{elpblack}{gray}{0.00}
\definecolor{elpgraydark}{gray}{0.25}
\definecolor{elpgraymid}{gray}{0.55}
\definecolor{elpgraylight}{gray}{0.96}
\RequirePackage[hyperfootnotes=false]{hyperref}
\RequirePackage[nameinlink,noabbrev]{cleveref}
\RequirePackage{titlesec}
\RequirePackage{fancyhdr}
\RequirePackage{setspace}
\RequirePackage{caption}
\RequirePackage{subcaption}
\RequirePackage{url}
\RequirePackage{float}
\RequirePackage{needspace}
\RequirePackage{tikz}
\usetikzlibrary{arrows.meta,positioning,calc,fit,backgrounds,decorations.pathreplacing,patterns}
\RequirePackage[
 backend=biber,style=authoryear,natbib=true,maxcitenames=2,maxbibnames=20,
 uniquename=false,doi=true,url=true,isbn=false,giveninits=true,dashed=false
]{biblatex}
\defbibheading{bibliography}[\refname]{\section*{#1}}
\hypersetup{hidelinks,pdfborder={0 0 0}}
\setlist{nosep,leftmargin=*}
\titleformat{\section}{\Needspace{7\baselineskip}\large\bfseries}{\thesection}{0.65em}{}
\titleformat{\subsection}{\Needspace{5\baselineskip}\normalsize\bfseries}{\thesubsection}{0.65em}{}
\titleformat{\subsubsection}{\normalsize\itshape}{\thesubsubsection}{0.65em}{}
\fancypagestyle{plain}{\fancyhf{}\fancyfoot[C]{\small\thepage}}
\newtheorem{theorem}{Theorem}[section]
\newtheorem{proposition}[theorem]{Proposition}

\theoremstyle{definition}
\newtheorem{definition}[theorem]{Definition}

\theoremstyle{remark}

\newcommand{\paperversion}[1]{\gdef\PaperVersion{#1}}
\newcommand{\paperseries}[1]{\gdef\PaperSeries{#1}}
\newcommand{\papersubtitle}[1]{\gdef\PaperSubtitle{#1}}
\newcommand{\paperaffiliation}[1]{\gdef\PaperAffiliation{#1}}
\newcommand{\papercontact}[1]{\gdef\PaperContact{#1}}

\AtBeginDocument{%
  \providecommand{\PaperVersion}{draft}%
  \providecommand{\PaperSeries}{ForesightFlow Research \textperiodcentered{} Event-Linked Perpetuals}%
  \providecommand{\PaperSubtitle}{}%
  \providecommand{\PaperAffiliation}{}%
  \providecommand{\PaperContact}{}%
  \providecommand{\PaperStatus}{}%
}
\newcommand{\authoritynote}[1]{\begin{center}\begin{minipage}{0.92\linewidth}\small\textit{#1}\end{minipage}\end{center}}
\tikzset{
 pbox/.style={draw=black,rounded corners=2pt,align=center,inner sep=5pt,font=\footnotesize,text width=23mm,minimum height=9mm,fill=white},
 pwide/.style={draw=black,rounded corners=2pt,align=center,inner sep=5pt,font=\footnotesize,text width=31mm,minimum height=10mm,fill=white},
 psoft/.style={draw=black,dashed,rounded corners=2pt,align=center,inner sep=5pt,font=\footnotesize,text width=26mm,minimum height=9mm,fill=white},
 parr/.style={-{Latex[length=2.0mm]},line width=0.65pt,draw=black},
 pdash/.style={-{Latex[length=2.0mm]},line width=0.65pt,dashed,draw=black},
 pbrace/.style={decorate,decoration={brace,amplitude=4pt},line width=0.55pt,draw=black},
 ffbox/.style={pbox},ffsmall/.style={pbox,font=\scriptsize,inner sep=4pt},ffsoft/.style={psoft},ffarrow/.style={parr},ffdashed/.style={pdash}
}
\makeatletter
\renewcommand{\maketitle}{%
  \begin{center}
    {\small\scshape \PaperSeries\par}
    \vspace{0.8em}
    {\LARGE\bfseries \@title\par}
    \ifx\PaperSubtitle\@empty\else\vspace{0.35em}{\large \PaperSubtitle\par}\fi
    \vspace{0.75em}
    {\large \@author\par}
    \ifx\PaperAffiliation\@empty\else\vspace{0.18em}{\small \PaperAffiliation\par}\fi
    \ifx\PaperContact\@empty\else\vspace{0.12em}{\small \PaperContact\par}\fi
    \vspace{0.32em}{\small \@date\par}
  \end{center}
  \ifx\PaperStatus\@empty\else\authoritynote{\PaperStatus}\fi
  \vspace{0.25em}
}
\makeatother

\paperseries{ForesightFlow Research \textperiodcentered{} Event-Linked Perpetuals \textperiodcentered{} Paper 7.4}
\paperversion{r0.9.3}
\papersubtitle{Terminal-Value Gaps, Trading Availability, and Venue Finality on Kalshi}
\title{Price Discovery at the Boundary of Contractual Decidability}
\author{Maksym Nechepurenko}
\paperaffiliation{Research Department of Devnull FZCO, Dubai, UAE}
\papercontact{maksym@devnull.ae \quad ORCID 0000-0002-9515-8841}
\date{September 2026 \textbar{} Version r0.9.3}
\hypersetup{pdftitle={Price Discovery at the Boundary of Contractual Decidability: Terminal-Value Gaps, Trading Availability, and Venue Finality on Kalshi},pdfauthor={Maksym Nechepurenko},pdfsubject={Event-Linked Perpetuals --- Kalshi Research Track},pdfkeywords={prediction markets, price discovery, contractual decidability, market closure, Kalshi}}
\begin{document}
\maketitle
\begingroup
\renewcommand{\thefootnote}{}
\footnotetext{\footnotesize\textit{ELP --- Kalshi Research Track.} This paper forms part of the Kalshi-focused research track within the Event-Linked Perpetuals series.}
\endgroup
\begingroup\footnotesize
\begin{abstract}
This paper studies price discovery around contractual decidability rather than an arbitrary venue label. Its upstream lifecycle and decidability clocks are specified in Papers~7.1 and~7.3 \parencite{elp71preprint,elp73preprint}. Historical venue endpoints remain useful background: 152,694 ordinary markets form the retrospective feasibility denominator, 71,657 have an exact public endpoint, and 70,979 have an exact determination-to-endpoint pair. Those fields do not supply a contractual-decidability clock.

The completed historical recovery produced no historically admissible contractual-decidability cohort. The prospective infrastructure shakedown has passed and evidence enrollment is active, but production price extraction has not started. The primary binary cohort will be drawn from the prospectively enrolled and blind-adjudicated contractual-decidability frame, with an accepted exact or interval first-decidability clock, exact terminal payoff, trading-availability classification, and admissible bounded non-block trade or real-candle coverage.

Markets tradable after decidability enter a reaction cohort; markets closed before decidability enter a stale-terminal cohort. Closure is a competing event for convergence, not ordinary missingness. A 20-market price pilot may validate acquisition, block-trade treatment, synthetic-candle rejection, staleness, and clock alignment only after blind packet lock and only from accepted prospective clock candidates. It does not create price estimates.

The paper specifies a prospective primary cohort and a targeted pilot protocol. Price estimation awaits independently reconstructed first/stable-decidability clocks and source-release clusters. No broad exchange-wide trade crawl or post hoc clock substitution is permitted.
\end{abstract}
\endgroup
\noindent\textbf{Keywords:} prediction markets; price discovery; contractual decidability; market closure; Kalshi.\\
\textbf{JEL:} G13, G14, G18.

\section{Introduction}
Prediction-market prices are often read as probabilities, but their terminal boundary is also a funding and market-access problem. A winning claim that pays one dollar later need not trade at one dollar today even after outcome uncertainty is small. Capital remains tied up; settlement can be contested or delayed; and sparse trading can leave an old price as the final visible observation. Recent evidence on settlement discounting shows why raw near-certainty gaps cannot be interpreted as forecast error alone \parencite{gebele2026settlementdiscount}.

This paper studies the boundary on Kalshi with five clocks kept distinct: source publication, first contractual decidability, market close, first venue determination, and finalization. A determination-anchored event study alone can miss an earlier response by classifying it as pre-event movement. Conversely, a market closed before decisive evidence cannot display a post-event transaction response at all. The observation design must therefore precede price estimation.

The paper complements broad work on price discovery across modern prediction venues \parencite{ng2026pricediscovery} and classical event-study methodology \parencite{mackinlay1997event}. It does not estimate manipulation, though settlement-source manipulation is an adjacent mechanism \parencite{dai2026settlement}. It asks how observed prices approach terminal value around independently constructed institutional clocks.

Kalshi studies based on time-to-close, final traded prices, or macroeconomic announcement windows establish that its prices contain information and exhibit nontrivial microstructure \parencite{burgi2026makers,diercks2026macro}. Paper 7.4 does not repeat that exercise. Its novelty is conditional event time: the reaction clock is constructed independently from rules and public evidence, and market closure is treated as an economic competing event rather than a missing post-event price.

\section{Cohort and data surfaces}
The primary cohort consists of markets selected from the prospectively enrolled and blind-adjudicated contractual-decidability frame. The frame must establish an accepted exact or interval first-decidability clock, exact terminal payoff, trading-availability classification, and bounded admissible non-block trade or real-candle coverage. Markets are then partitioned before analysis:
\begin{itemize}
\item \emph{Reaction-eligible}: trading is available and admissible prices exist on both sides of decidability;
\item \emph{Close competing}: trading closes after decidability but before price convergence;
\item \emph{Stale terminal}: trading closed before decidability;
\item \emph{Interval-valued clock}: decidability is used only for bounds or sensitivity;
\item \emph{Price unobserved}: no admissible price falls within the staleness rule.
\end{itemize}
The retrospective recovery programme did not produce a historically admissible contractual-decidability cohort. Historical determination-anchored or terminal-price summaries, if used, remain separate descriptive diagnostics; they cannot be substituted into the primary decidability-anchored event study.

The primary data surface is the public trade record, which provides market identity, yes/no dollar prices, quantity, time, and a block-trade flag \parencite{kalshiGetTrades2026,kalshiHistoricalTrades2026}. Primary event windows exclude block trades; they enter a labelled sensitivity. One-minute candles are secondary. Kalshi's continuity option can prepend a synthetic candle with null OHLC values and a carried previous price; such records are never treated as observed trades or real candles \parencite{kalshiGetCandles2026,kalshiHistoricalCandles2026}.

The prospective infrastructure shakedown passed and prospective enrollment began on 4 September 2026 at 22:54:26 UTC. Its price layer remained empty throughout the shakedown; production price extraction has not started. The upstream programme preserves prospective rule and official source-release evidence, but neither price support nor price results are reported in this paper.

Registered horizons are 24 hours, 6 hours, 1 hour, 10 minutes, and 1 minute before the clock, with 1 minute, 10 minutes, 1 hour, and 2 hours after where trading remains available. The extraction grid is specified before analysis.

\begin{figure}[H]
\centering
\begin{tikzpicture}[x=1cm,y=1cm]
\node[pbox,text width=26mm] (src) at (-4.9,1.1) {Source\\publication};
\node[pbox,text width=26mm] (dec) at (-1.7,1.1) {First\\decidability};
\node[pbox,text width=26mm] (close) at (1.7,1.1) {Market\\close};
\node[pbox,text width=28mm] (det) at (-0.1,-1.25) {Venue\\determination};
\node[pbox,text width=25mm] (fin) at (3.25,-1.25) {Finalization};
\draw[parr] (det.east)--(fin.west);
\draw[pdash] (src.south) -- ++(0,-0.38) -| (det.north west);
\draw[pdash] (dec.south) -- (det.north);
\draw[pdash] (close.south) -- ++(0,-0.38) -| (det.north east);
\end{tikzpicture}
\caption{Partial clock order for the registered design. Source publication, first decidability, and market close are analytical coordinates whose relative order is classified market by market. Only venue determination to finalization is shown as a fixed venue-side order; dashed links do not assert a universal pre-determination sequence or a post-decidability trading interval.}
\label{fig:74-clocks}
\end{figure}
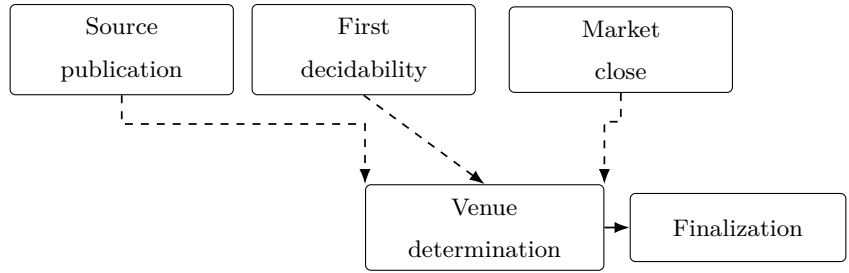

\newpage
\section{Prospective activation and acquisition prerequisites}\label{sec:74-feasibility}
The historical recovery establishes venue-side background and the prospective shakedown establishes observation capability. The relevant primary price-study population is not the full KMVE object universe or the historical ordinary-market frame. It is the much smaller intersection of:
\begin{enumerate}
  \item a market selected from the prospectively enrolled Paper 7.3 frame;
  \item an exact terminal payoff and endpoint by the registered follow-up boundary;
  \item an accepted exact or interval first-decidability clock from blind Paper 7.3 adjudication;
  \item exact trading-availability classification and bounded admissible price coverage.
\end{enumerate}

\begin{table}[H]
\centering\footnotesize
\caption{Inputs required for the Paper 7.4 study design.}
\begin{tabularx}{\textwidth}{@{}L{0.29\textwidth}L{0.27\textwidth}Y@{}}
\toprule
Input & Current state & Consequence \\
\midrule
Exact market and event identities & Available from the registered population and ordinary-market key registry & Deterministic linkage; no title or fuzzy matching \\
\addlinespace[2pt]
Historical endpoint background & 71,657 exact endpoint; 81,037 no exact endpoint among 152,694 ordinary markets & Descriptive anchor sensitivity only; not the primary cohort \\
\addlinespace[2pt]
Prospective evidence enrollment & Active after a passed 72-hour shakedown & Primary cohort is formed only at enrollment close \\
\addlinespace[2pt]
First/stable decidability & Accepted blind-adjudication clock required & No primary event anchor until the clock is available \\
\addlinespace[2pt]
Trade and real-candle extraction & Inactive during the shakedown & Retrieve only after accepted blind clocks exist \\
\addlinespace[2pt]
Block-trade and synthetic-candle treatment & Defined in the acquisition protocol & Publication windows can be audited \\
\addlinespace[2pt]
Independent source-release clusters & Independently reconstructed clusters required & Effective sample size and dependence units remain unknown \\
\bottomrule
\end{tabularx}
\label{tab:74-activation}
\end{table}

\subsection{Historical background and prospective registry}
The historical venue-side registry has four load-bearing classes:
\begin{itemize}
  \item 70,979 markets with an exact determination-to-endpoint pair;
  \item 678 markets with an exact endpoint but without the paired determination clock required for determination-anchor sensitivity;
  \item 81,037 markets retained in attrition and censoring summaries but excluded from terminal-value-gap estimation because no exact endpoint is observed by the follow-up boundary;
  \item eight ordinary-cohort tickers whose registered interval intersects a documented gap and are retained with coordinate-specific coverage grades.
\end{itemize}
These classes are not a price cohort. They support lifecycle eligibility and descriptive anchor sensitivity, but do not identify a historical contractual-decidability event study. The primary registry is formed only after exact intersection with the prospectively enrolled Paper 7.3 frame and accepted blind clocks. Endpoint evidence identifies terminal payoff; it does not reveal historical price availability or source-publication time.

\subsection{No broad price crawl}
The KMVE V1 population contains 7,611,594 MVE market objects \parencite{nechepurenko2026kmve}. Downloading trade or candle histories for all of them would not answer the study question and would create pseudoreplication across related contracts. Acquisition is restricted to selected Paper 7.3 candidates with independently reconstructed clocks and compatible lifecycle states. One source release, event, or collection remains the dependence unit even when it governs many tickers.

\subsection{Twenty-market pilot}
A limited pilot may use up to 20 markets drawn from accepted prospective clock candidates after blind packet lock. It verifies acquisition availability, price transformation, block-trade treatment, synthetic-candle rejection, price-age and staleness rules, and source/decidability/close/determination alignment. It reports only acquisition and coverage diagnostics; it does not estimate an average reaction and is not expanded after inspecting price paths.

\subsection{Acquisition prerequisites}
Production extraction begins only after accepted blind prospective clocks exist. A market lacking an accepted first-decidability clock with a source-release cluster, an exact terminal payoff, trading-availability classification, or bounded admissible price coverage is outside the primary event-study cohort. Venue determination is never substituted for decidability, and endpoint-only historical cases do not enter the primary study.

\newpage
\section{Outcome-aligned price and terminal-value gap}
For binary market $i$, let $y_i=1$ when YES wins and $y_i=0$ when NO wins. Let $p_i^Y(t)$ and $p_i^N(t)$ be the admissible transaction prices. Define the outcome-aligned winning-side price
\[
W_i(t)=y_i p_i^Y(t)+(1-y_i)p_i^N(t),
\]
and the raw terminal-value gap
\[
g_i(t)=1-W_i(t).
\]
The term ``gap'' is deliberate. It is not assumed to equal forecast error or disagreement.

Every price carries age $a_i(t)=t-t_i^{last}$. An observation older than the registered staleness threshold is missing for reaction estimation but retained for the stale-terminal analysis. Scalar contracts require a separate payoff mapping and are excluded from the primary binary estimand.

To display the possible funding component, define a sensitivity benchmark for annualized rate $r$:
\[
b_i(t;r)=\exp\{-r(t_{final,i}-t)/365d\},\qquad
q_i(t;r)=b_i(t;r)-W_i(t).
\]
This is not asserted to be an arbitrage-free fair value; it omits fees, contestability, credit, and execution. It asks how much of the raw gap remains after a transparent time-value adjustment across a grid of $r$.

\section{Admissibility and competing events}
\begin{definition}[Reaction-eligible market]
A market is reaction-eligible around $t^{first}_{decide}$ when the lifecycle ledger establishes trading availability and at least one admissible non-block transaction exists in registered pre- and post-clock windows.
\end{definition}

\begin{definition}[Stale-terminal market]
A market is stale-terminal when trading closes before first decidability. It supports a last-price-age estimand but not an intramarket post-decidability response.
\end{definition}

\begin{proposition}[Determination anchoring can attenuate measured reaction]
If decisive public evidence precedes venue determination and prices respond before $t_{det}$, a determination-anchored study mechanically allocates part or all of the response to the pre-event period.
\end{proposition}

\begin{proposition}[Closure creates structural missingness]
When trading is unavailable after decidability, absence of post-clock transactions is not evidence that beliefs failed to update. It is absence of the price observation process.
\end{proposition}

\begin{figure}[H]
\centering
\begin{tikzpicture}[x=1cm,y=1cm]
\node[ffsmall,text width=42mm](q) at (0,0){Trading available\\after first decidability?};
\node[ffsoft,text width=37mm](yes) at (-3.0,-1.7){Reaction cohort\\gap change and convergence};
\node[ffsoft,text width=37mm](no) at (3.0,-1.7){Stale-terminal cohort\\last gap and price age};
\draw[ffarrow](q)--node[left,font=\scriptsize]{yes}(yes);
\draw[ffarrow](q)--node[right,font=\scriptsize]{no}(no);
\end{tikzpicture}
\caption{Cohort admissibility. Closure before or after decidability is handled as a separate economic event, not as ordinary missing data.}
\label{fig:74-admissibility}
\end{figure}
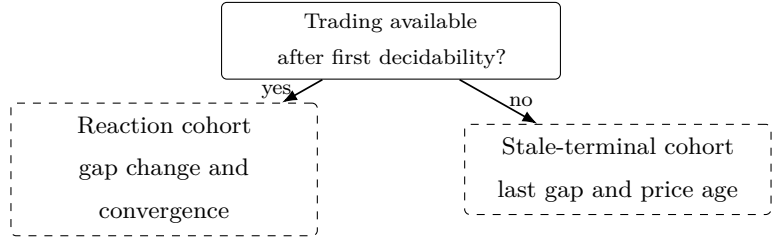

From first decidability, define three mutually exclusive outcomes: enter an $\epsilon$-neighborhood $g_i(t)\le\epsilon$ while active; close before convergence; or remain active without convergence until administrative censoring. Cumulative incidence is estimated with Aalen--Johansen rather than treating closure as independent right censoring \parencite{aalen1978transition,andersen1993counting}.

\section{Primary estimands}
For reaction-eligible markets and clock $\tau$ define gap reduction
\[
\Delta g_i(h_-,h_+;\tau)=g_i(\tau-h_-)-g_i(\tau+h_+).
\]
Positive values indicate movement toward terminal value. The time to convergence is
\[
T_i(\epsilon)=\inf\{t\ge t^{first}_{decide,i}:g_i(t)\le\epsilon\}-t^{first}_{decide,i},
\]
with closure as a competing event.

For stale-terminal markets let
\[
g_i^{last}=g_i(t_i^{last}),\qquad A_i^{last}=t^{first}_{decide,i}-t_i^{last}.
\]
The pair measures both terminal-value distance and how old the last tradable price was when the public contract first became decidable.

Anchor sensitivity compares the same markets around $t_{source}$, $t^{first}_{decide}$, and $t_{det}$. Interval-decidability cases produce event-time bounds or a separate sensitivity; no midpoint enters the primary study.

\begin{table}[H]
\centering\footnotesize
\caption{Primary analyses and dependence units.}
\begin{tabularx}{\textwidth}{@{}p{0.23\textwidth}p{0.28\textwidth}X@{}}
\toprule
Analysis & Primary object & Dependence unit \\
\midrule
Reaction path & $g(t)$ and $\Delta g$ around first decidability & source release; event sensitivity \\
Convergence & $T(\epsilon)$ with closure competing & source release \\
Stale terminal & $(g^{last},A^{last})$ & event/series \\
Anchor sensitivity & source vs decidability vs determination & same-market paired; source cluster \\
Funding sensitivity & $q(t;r)$ over registered $r$ grid & event/source \\
\bottomrule
\end{tabularx}
\end{table}

\section{Registered empirical tests}
\textbf{T1 --- Pre-determination incorporation.} Among reaction-eligible markets, compare gap paths before first decidability and before venue determination. The estimand is descriptive convergence, not causal efficiency.

\textbf{T2 --- Anchor displacement.} Quantify how the apparent event-time response changes when $t_{det}$ replaces $t^{first}_{decide}$ for the same markets.

\textbf{T3 --- Competing closure.} Estimate cumulative incidence of convergence versus market closure after first decidability. Report support at every event-time bin.

\textbf{T4 --- Stale-terminal exposure.} Describe $g^{last}$ jointly with price age for markets closed before first decidability; do not pool them with reaction-eligible markets.

\textbf{T5 --- Settlement-discount sensitivity.} Recompute terminal-value gaps relative to $b_i(t;r)$ across a registered rate grid. Remaining residuals are not labeled mispricing without execution and risk analysis.

\section{Inference and robustness}
Event-time summaries report median, quantiles, and the number of distinct markets, events, and source releases contributing at each bin. Cluster bootstrap uses source release as primary unit and event/series as sensitivities. A large number of tickers sharing one publication is not treated as equally many independent information shocks.

Robustness analyses include: matched pseudo-event clocks; source, decidability, and determination anchor substitution; multiple staleness thresholds; non-block versus all trades; trade-level versus real-candle measures; exclusion of reconnect-affected lifecycle intervals; and exact-clock-only analysis. Synthetic continuity candles are excluded in every specification. A pooled regression, if shown, is secondary to raw support and event-time distributions.

\newpage
\section{Validity and reproducibility}
The price study requires cohort-linked trade observations, documented endpoint semantics, exact timestamp precision, block-trade status, clock provenance, trading-availability state, and reproducible reconstruction. The Paper 7.1 and 7.3 registries provide the lifecycle and decidability inputs; they cannot be recreated from prices alone. Historical observations lacking rule history cannot enter the primary decidability-anchored cohort.

The result order is fixed: cohort/clock attrition; price-data coverage; clock ordering; pre-event gaps; reaction paths; competing closure; stale-terminal analysis; financing sensitivity; negative controls.

\section{Limitations}
Transaction prices are not executable prices for arbitrary size. Sparse trading, fees, and bid--ask spreads remain relevant. Conditioning on final settlement selects resolved markets; unresolved markets appear in attrition but not in terminal-payoff estimands. Exact-decidability analysis will be restricted to the subset with accepted exact clocks; interval and unmeasured cases remain separate. No accepted prospective decidability clocks exist yet. Funding sensitivity is not a valuation model. The design does not identify private information, trader identity, manipulation, or causal effects of venue policy.

\section{Conclusion}
The historical recovery did not produce a historically admissible contractual-decidability cohort. Venue-side historical endpoints support lifecycle eligibility and descriptive anchor sensitivity, but they do not identify a historical contractual-decidability event study. The primary price analysis therefore depends on prospectively preserved rule, source-release, and decidability evidence.

The prospective infrastructure shakedown has passed and enrollment is active, but production price extraction has not started and no price estimate exists. The event anchor remains contractual decidability, reconstructed independently from rules and public release objects. Accepted prospective clocks remain a dependency for the price protocol. Bounded trade and real-candle acquisition is restricted to selected candidates only after accepted blind first/stable clocks and source-release clusters exist, with closure treated as a competing event and stale-terminal markets analyzed separately.

The design separates historical venue-side background from the prospective contractual-decidability frame. Keeping those roles separate prevents a convenient determination timestamp, a terminal snapshot, or a large market-object population from substituting for the scientific event boundary.

\appendix
\newpage
\section{Price-provenance rules}
Every observation window records its endpoint, parameters, market identifier, retrieval time, timestamp precision, block-trade flag, and price-field semantics. Derived candles and resampled series remain separate from raw trades. Any carried or synthetic candle is flagged and excluded from primary observation counts.

\section*{Data and Code Availability}
The population and hierarchy layer is available as the \emph{Kalshi Multivariate Event Market Materialization Dataset (KMVE)}, Mendeley Data, V1, doi: \href{https://doi.org/10.17632/fn65786cg6.1}{10.17632/fn65786cg6.1}. No price-result dataset exists, no production-enrollment price data are claimed or distributed, and no key-level candidate registry, bounded price response, or raw trade response accompanies this preprint. The paper specifies a bounded prospective protocol; it does not report price estimates or accepted prospective clocks.

\section*{Generative AI Disclosure}
OpenAI ChatGPT and Codex were used for editorial and technical assistance during manuscript preparation. The author made all substantive research decisions, reviewed the final manuscript, and assumes full responsibility for its contents.

\section*{Funding}
This research received no external funding.

\section*{Competing Interests}
The author is affiliated with the Research Department of Devnull FZCO and leads the ForesightFlow research programme. No external sponsor influenced the research design, analysis, interpretation, or decision to publish. The article does not evaluate a commercial product or make investment recommendations.

\printbibliography[heading=bibliography,title={References}]

@article{aalen1978transition,
  author={Aalen, Odd O. and Johansen, S{\o}ren},
  title={An Empirical Transition Matrix for Non-Homogeneous Markov Chains Based on Censored Observations},
  journal={Scandinavian Journal of Statistics}, year={1978}, volume={5}, number={3}, pages={141--150}}

@book{andersen1993counting,
  author={Andersen, Per Kragh and Borgan, {\O}rnulf and Gill, Richard D. and Keiding, Niels},
  title={Statistical Models Based on Counting Processes}, publisher={Springer}, location={New York},
  year={1993}, doi={10.1007/978-1-4612-4348-9}}

@article{mackinlay1997event,
  author={MacKinlay, A. Craig}, title={Event Studies in Economics and Finance},
  journal={Journal of Economic Literature}, year={1997}, volume={35}, number={1}, pages={13--39}}

@online{ng2026pricediscovery,
  author={Ng, Hunter and Peng, Lin and Tao, Yubo and Zhou, Dexin},
  title={Price Discovery and Trading in Modern Prediction Markets}, year={2026},
  note={Preprint, SSRN 5331995}, doi={10.2139/ssrn.5331995}, url={https://doi.org/10.2139/ssrn.5331995}}

@article{dai2026settlement,
  author={Dai, David and Jia, Ruizhe and Yu, Shihao}, title={Settlement Manipulation in Prediction Markets},
  year={2026}, eprint={2606.31675}, archivePrefix={arXiv}, primaryClass={q-fin.TR},
  url={https://arxiv.org/abs/2606.31675}}

@article{gebele2026settlementdiscount,
  author={Gebele, Jonas and Matthes, Florian},
  title={When Certainty Is Not Worth It: Capital Lock-Up and Settlement Discounting in Prediction Markets},
  year={2026}, eprint={2605.31431}, archivePrefix={arXiv}, primaryClass={q-fin.TR},
  url={https://arxiv.org/abs/2605.31431}}

@online{kalshiGetTrades2026, author={{Kalshi}}, title={Get Trades}, year={2026},
  url={https://docs.kalshi.com/api-reference/market/get-trades}, urldate={2026-08-15}, note={Kalshi Exchange API reference}}

@online{kalshiGetCandles2026, author={{Kalshi}}, title={Get Market Candlesticks}, year={2026},
  url={https://docs.kalshi.com/api-reference/market/get-market-candlesticks}, urldate={2026-08-15}, note={Kalshi Exchange API reference}}

@online{kalshiHistoricalCandles2026, author={{Kalshi}}, title={Get Historical Market Candlesticks}, year={2026},
  url={https://docs.kalshi.com/api-reference/historical/get-historical-market-candlesticks}, urldate={2026-08-15}, note={Kalshi Exchange API reference}}

@online{kalshiHistoricalTrades2026, author={{Kalshi}}, title={Get Historical Trades}, year={2026},
  url={https://docs.kalshi.com/api-reference/historical/get-historical-trades}, urldate={2026-08-15}, note={Kalshi Exchange API reference}}

@report{burgi2026makers,
  author={B{\"u}rgi, Constantin and Deng, Wanying and Whelan, Karl},
  title={Makers or Takers: The Economics of the Kalshi Prediction Market},
  institution={Center for Economic Research, George Washington University},
  number={2026-001}, year={2026},
  url={https://www2.gwu.edu/~forcpgm/2026-001.pdf}, urldate={2026-08-15}}

@report{diercks2026macro,
  author={Diercks, Anthony M. and Katz, Jared Dean and Wright, Jonathan H.},
  title={Kalshi and the Rise of Macro Markets},
  institution={Board of Governors of the Federal Reserve System},
  series={Finance and Economics Discussion Series}, number={2026-010}, year={2026},
  doi={10.17016/FEDS.2026.010}}

@dataset{nechepurenko2026kmve,
  author    = {Nechepurenko, Maksym},
  title     = {Kalshi Multivariate Event Market Materialization Dataset (KMVE)},
  publisher = {Mendeley Data},
  year      = {2026},
  version   = {V1},
  doi       = {10.17632/fn65786cg6.1},
  url       = {https://doi.org/10.17632/fn65786cg6.1}
}

@online{elp71preprint,
  author = {Nechepurenko, Maksym},
  title = {Outcome Determination and Settlement Finality on Kalshi: Public State Paths, Prospective Measurement, and Empirical Identification},
  date = {2026-09-01},
  doi = {10.2139/ssrn.7421258},
  url = {https://ssrn.com/abstract=7421258},
  note = {SSRN 7421258}
}

@online{elp73preprint,
  author = {Nechepurenko, Maksym},
  title = {From Public Evidence to Contractual Outcome: First and Stable Decidability on Kalshi},
  date = {2026-09-03},
  doi = {10.2139/ssrn.7421261},
  url = {https://ssrn.com/abstract=7421261},
  note = {SSRN 7421261}
}
\end{document}